\documentclass[preprint,showpacs,preprintnumbers,amsmath,amssymb,superscriptaddress]{revtex4}
\usepackage{CJK}
\usepackage{graphicx}% Include figure files
\usepackage{dcolumn}% Align table columns on decimal point
\usepackage{bm}% bold math
\usepackage{tabularx}

\begin{document}
%\begin{CJK*}{GBK}{song}
%\preprint{APS/123-QED}

\title{Proton widths of the low-lying $^{16}$F states from the $^{15}$N($^{7}$Li,\,$^{6}$Li)$^{16}$N
reaction}% Force line breaks with \\

\author{Z. D. Wu}\affiliation{China Institute of Atomic Energy, P.O. Box 275(10),
Beijing 102413, China}\affiliation{Institute of Modern Physics, Shanxi Normal University,
Linfen 041004, China}
\author{B.
Guo}\thanks{Corresponding author:
guobing@ciae.ac.cn}\affiliation{China Institute of Atomic Energy,
P.O. Box 275(10), Beijing 102413, China}
\author{Z. H. Li}\affiliation{China Institute of Atomic Energy, P.O. Box 275(10),
Beijing 102413, China}
\author{Y. J. Li}\affiliation{China Institute of Atomic Energy, P.O. Box 275(10),
Beijing 102413, China}
\author{J. Su}\affiliation{China Institute of Atomic Energy, P.O. Box 275(10),
Beijing 102413, China}
\author{D. Y.
Pang}\affiliation{School of Physics and Nuclear Energy
Engineering, Beihang University, Beijing 100191,
China}\affiliation{International Research Center for Nuclei and
Particles in the Cosmos, Beihang University, Beijing 100191,
China}
\author{S. Q. Yan}\affiliation{China Institute of Atomic Energy, P.O. Box 275(10),
Beijing 102413, China}
\author{E. T.
Li}\affiliation{College of Physics Science and Technology,
Shenzhen University, Shenzhen 518060, China}
\author{X. X. Bai}\affiliation{China Institute of Atomic Energy, P.O. Box 275(10),
Beijing 102413, China}
\author{X. C. Du}\affiliation{China Institute of Atomic Energy, P.O. Box 275(10),
Beijing 102413, China}
\author{Q. W. Fan}\affiliation{China Institute of Atomic Energy, P.O. Box 275(10),
Beijing 102413, China}
\author{L. Gan}\affiliation{China Institute of Atomic Energy, P.O. Box 275(10),
Beijing 102413, China}
\author{J. J.
He}\affiliation{Institute of Modern Physics, Chinese Academy of
Sciences (CAS), Lanzhou 730000, China}
\author{S. J. Jin}\affiliation{China Institute of Atomic Energy, P.O. Box 275(10),
Beijing 102413, China}
\author{L. Jing}\affiliation{China Institute of Atomic Energy, P.O. Box 275(10),
Beijing 102413, China}
\author{L. Li}\affiliation{Institute of Modern Physics, Chinese Academy of
Sciences (CAS), Lanzhou 730000, China}
\author{Z. C. Li}\affiliation{China Institute of Atomic Energy, P.O. Box 275(10),
Beijing 102413, China}
\author{G. Lian}\affiliation{China Institute of Atomic Energy, P.O. Box 275(10),
Beijing 102413, China}
\author{J. C. Liu}\affiliation{China Institute of Atomic Energy, P.O. Box 275(10),
Beijing 102413, China}
\author{Y. P. Shen}\affiliation{China Institute of Atomic Energy, P.O. Box 275(10),
Beijing 102413, China}
\author{Y. B. Wang}\affiliation{China Institute of Atomic Energy, P.O. Box 275(10),
Beijing 102413, China}
\author{X. Q. Yu}\affiliation{Institute of Modern Physics, Chinese Academy of
Sciences (CAS), Lanzhou 730000, China}
\author{S. Zeng}\affiliation{China Institute of Atomic Energy, P.O. Box 275(10),
Beijing 102413, China}
\author{D. H. Zhang}\affiliation{Institute of Modern Physics, Shanxi Normal University,
Linfen 041004, China}
\author{L. Y. Zhang}\affiliation{Institute of Modern Physics, Chinese Academy of
Sciences (CAS), Lanzhou 730000, China}
\author{W. J. Zhang}\affiliation{China Institute of Atomic Energy, P.O. Box 275(10),
Beijing 102413, China}
\author{W. P. Liu}\affiliation{China Institute of Atomic Energy, P.O. Box 275(10),
Beijing 102413, China}

\date{\today}% It is always \today, today,

\begin{abstract}

All the $^{16}$F levels are unbound by proton emission. To date
the four low-lying $^{16}$F levels below 1 MeV have been
experimentally identified with well established spin-parity values
and excitation energies with an accuracy of 4\,-\,6 keV. However,
there are still considerable discrepancies for their level widths.
The present work aims to explore these level widths through an
independent method. The angular distributions of the
$^{15}$N($^{7}$Li,\,$^{6}$Li)$^{16}$N reaction leading to the
first four states in $^{16}$N were measured using a high-precision
Q3D magnetic spectrograph. The neutron spectroscopic factors and
the asymptotic normalization coefficients for these states in
$^{16}$N were then derived based on distorted wave Born
approximation analysis. The proton widths of the four low-lying
resonant states in $^{16}$F were obtained according to charge
symmetry of strong interaction.

\end{abstract}

\maketitle

\section{Introduction}

\begin{figure}[htbp]
\begin{center}
\resizebox{0.45\textwidth}{!}{
  \includegraphics{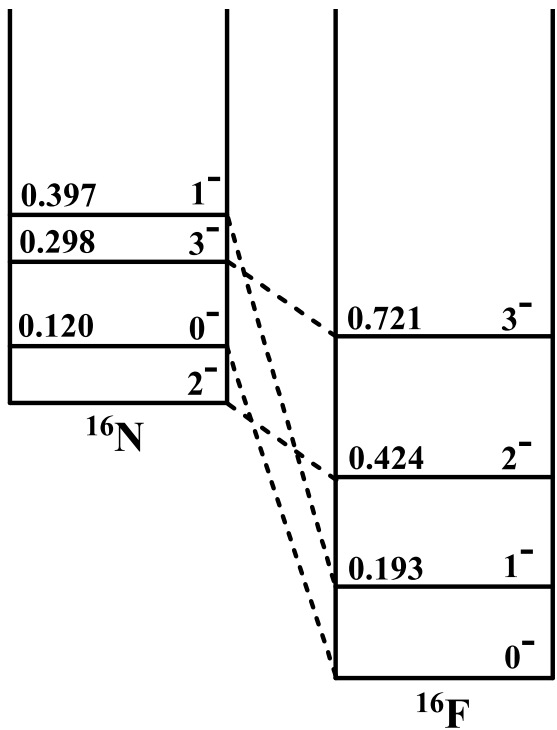}}
\caption{Level diagram for the low-lying states in mirror pair of
$^{16}$N-$^{16}$F.} \label{fig1}
\end{center}
\end{figure}

In the past there has been considerable effort to explore the
structure of $^{16}$N, while there is fewer report for its mirror
analog $^{16}$F since it can be investigated through relatively
few reactions including $^{14}$N($^3$He,\,$n$)$^{16}$F
\cite{zaf65,boh73,ots76}, $^{16}$O($p$,\,$n$)$^{16}$F
\cite{mos71,faz82,ori82,ohn87,mad97},
$^{16}$O($^3$He,\,$t$)$^{16}$F \cite{peh65,nan77,ste84,fuj09}, and
$^{19}$F($^{3}$He,\,$^{6}$He)$^{16}$F \cite{nan77}. The level
diagram for the four low-lying states in mirror pair of
$^{16}$N-$^{16}$F is shown in Fig. \ref{fig1}. All the states in
$^{16}$F are unbound and decay as $^{15}$O\,+\,$p$. The
measurements using stable beams have well determined spin-parity
values and excitation energies with an accuracy of 4\,-\,6 keV for
the four low-lying states in $^{16}$F \cite{til93}. However, these
measurements yielded only upper limits or rough estimates of the
$^{16}$F level widths. Recently, Lee et al. investigated the level
widths of these four states in $^{16}$F via the elastic resonance
scattering of $^{15}$O\,+\,$p$ based on a thick target inverse
kinematics method \cite{lee07}. Although these authors
significantly improved values for these level widths of $^{16}$F,
it is still desirable to perform a new measurement of these level
widths via an independent approach.

In the present work, we measure the angular distributions of the
$^{15}$N($^{7}$Li,\,$^{6}$Li)$^{16}$N reaction populating the four
low-lying states in $^{16}$N. The neutron spectroscopic factors
and the asymptotic normalization coefficients (ANCs) for these
states are then derived based on distorted wave Born approximation
(DWBA) analysis. The proton widths of the four low-lying resonant
states in the mirror analog $^{16}$F are extracted according to
charge symmetry of mirror nuclei. Similar approach has been
successfully used to study many mirror pairs such as
$^{12}$B-$^{12}$N \cite{guo07}, $^{15}$C-$^{15}$F \cite{tim06},
$^{27}$Mg-$^{27}$P \cite{guo06} and $^{57}$Ni-$^{57}$Cu
\cite{reh98}. Most recently, a short paper concerning the
$^{15}$N($^{7}$Li,\,$^{6}$Li)$^{16}$N angular distributions and
determination of the astrophysical
$^{15}$N($n$,\,$\gamma$)$^{16}$N reaction rate has been published
elsewhere \cite{guo14}.

\section{Experimental procedure}

The measurement of the angular distributions was performed at the
HI-13 tandem accelerator of the China Institute of Atomic Energy
(CIAE) in Beijing. The experimental setup and procedures are
similar to those reported previously \cite{guo12,li12,li13}. A
$^{7}$Li beam with an energy of 44 MeV was used to measure the
angular distributions of the $^{15}$N($^{7}$Li,\,$^{6}$Li)$^{16}$N
reaction populating the ground state and the first three excited
states at $E_x$ = 0.120, 0.298, and 0.397 MeV in $^{16}$N. In
addition, the angular distribution of the $^{7}$Li\,+\,$^{15}$N
elastic scattering was measured to obtain the optical model
potential (OMP) parameters for the entrance channel of the
transfer reaction. To extract the exit channel OMP parameters a
34.5 MeV $^6$Li beam was also delivered for the measurement of the
angular distribution for the $^6$Li\,+\,$^{15}$N elastic
scattering.

Melamine C$_3$N$_3$($^{15}$NH$_2$)$_3$ enriched to 99.35\% in
$^{15}$N was employed as target material with a thickness of 46
$\mu$g/cm$^2$, which was evaporated on a 30 $\mu$g/cm$^2$ thick
carbon foil. In addition, a $^{14}$N target was used for
background evaluation. To improve the thermal conductivity of the
targets a 22 $\mu$g/cm$^2$ thick gold was evaporated on melamine
foil. The target thickness was determined using an analytical
balance with a precision of 1 $\mu$g and was verified with the
well-known differential cross sections of the
$^{7}$Li\,+\,$^{15}$N elastic scattering at $\theta_\mathrm{c.m.}$
= 33.5$^\circ$ and 49.2$^\circ$ \cite{woo82,oli96}. After
considering the balance precision and the error of the
differential cross sections, an uncertainty of 5\% was assigned
for target thickness.

A movable Faraday cup covering an angular range of
$\pm\,$6$^\circ$ in laboratory frame was used to measure the beam
current for normalization of the cross sections at
$\theta_\mathrm{lab}\,>\,6^\circ$. The Faraday cup was removed
when measuring the cross sections at
$\theta_\mathrm{lab}\,\leq\,6^\circ$. A silicon $\Delta E\,-\,E$
telescope located at $\theta_\mathrm{lab}\,=\,25^\circ$ was
employed for normalization of the cross sections at
$\theta_\mathrm{lab}\,\leq\,6^\circ$ by measuring the elastic
scattering of the incident ions on the targets. The reaction
products were analyzed with a Q3D magnetic spectrograph and were
recorded by a two-dimensional position-sensitive silicon detector
(PSSD, 50\,$\times$\,50 mm) placed at the focal plane of the
spectrograph. The two-dimensional position information from the
PSSD enabled the products emitted into the acceptable solid angle
to be recorded completely. The energy information from the PSSD
was used to remove the impurities with the same magnetic rigidity.

\begin{figure}[htbp]
\begin{center}
\resizebox{0.6\textwidth}{!}{
  \includegraphics{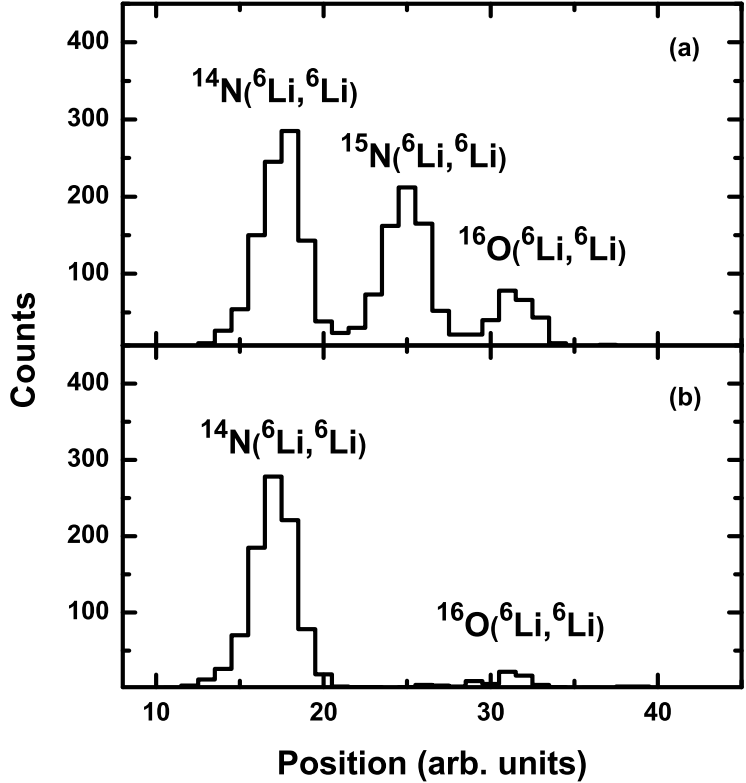}}
\caption{Focal-plane position spectra of the $^{6}$Li events at
$\theta_\mathrm{lab}$ = 18$^\circ$ from the elastic scattering on
the enriched $^{15}$N target (a) and the natural $^{14}$N target
(b).} \label{fig2}
\end{center}
\end{figure}

\begin{figure}[htbp]
\begin{center}
\resizebox{0.6\textwidth}{!}{
  \includegraphics{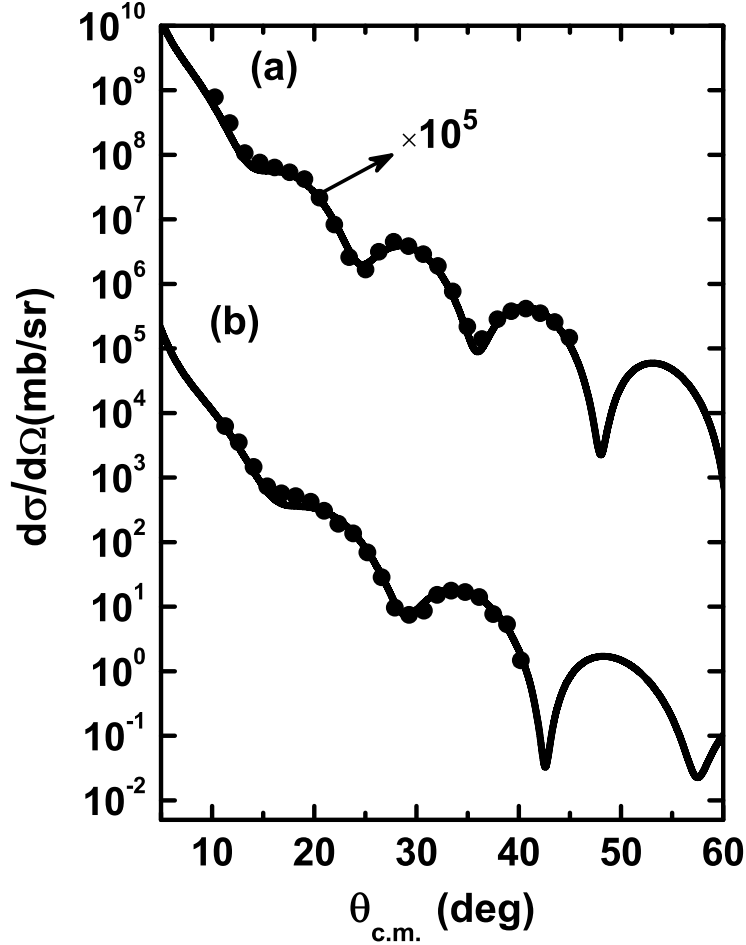}}
\caption{Angular distributions of the $^{7}$Li+$^{15}$N elastic
scattering at incident energy of 44 MeV (a) and the
$^{6}$Li+$^{15}$N elastic scattering at incident energy of 34.5
MeV (b). The solid curves represent the calculations with the
fitted OMP parameters.} \label{fig3}
\end{center}
\end{figure}

As an example, Figure \ref{fig2} displays the focal-plane position
spectra of the $^{6}$Li events at $\theta_\mathrm{lab}$ =
18$^\circ$ from the elastic scattering on the enriched $^{15}$N
target and the natural $^{14}$N target. One sees that the events
from the elastic scattering on different isotopes in the targets
can be clearly separated. The events from the elastic scattering
on carbon and gold ran out of the PSSD due to larger energy
differences. It should be mentioned that the elastic scattering
events from $^{15}$N and $^{14}$N cannot be separated any more
when measuring the cross sections at $\theta_\mathrm{lab}$ $<$
15$^\circ$. This is because the energy difference of $^{6}$Li from
the elastic scattering on different isotopes decreases with
$\theta_\mathrm{lab}$. Therefore, the background from $^{14}$N
needs to be evaluated to obtain the cross sections at
$\theta_\mathrm{lab}$ $<$ 15$^\circ$. The angular distributions of
the elastic scattering were obtained after background substraction
and beam normalization, as shown in Fig. \ref{fig3}.

\begin{figure}[htbp]
\begin{center}
\resizebox{0.6\textwidth}{!}{
  \includegraphics{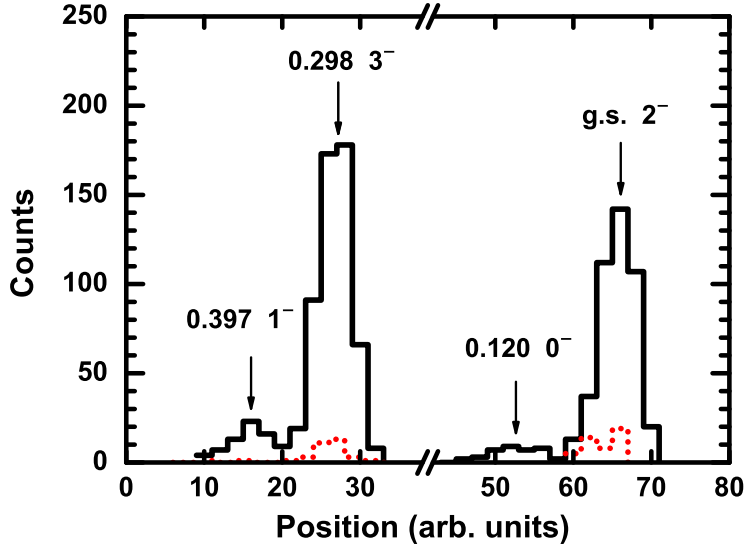}}
\caption{(Color online) Focal-plane position spectrum of the
$^{6}$Li events at $\theta_\mathrm{lab}$ = 10$^\circ$ from the
$^{15}$N($^{7}$Li,\,$^{6}$Li)$^{16}$N reaction. The black solid
and red dashed lines denote the results from the enriched $^{15}$N
target and the natural $^{14}$N target, respectively. The break in
the x-axis denotes the narrow gap between two separated
detectors.} \label{fig4}
\end{center}
\end{figure}

\begin{figure*}[htbp]
\begin{center}
\resizebox{0.9\textwidth}{!}{
\includegraphics{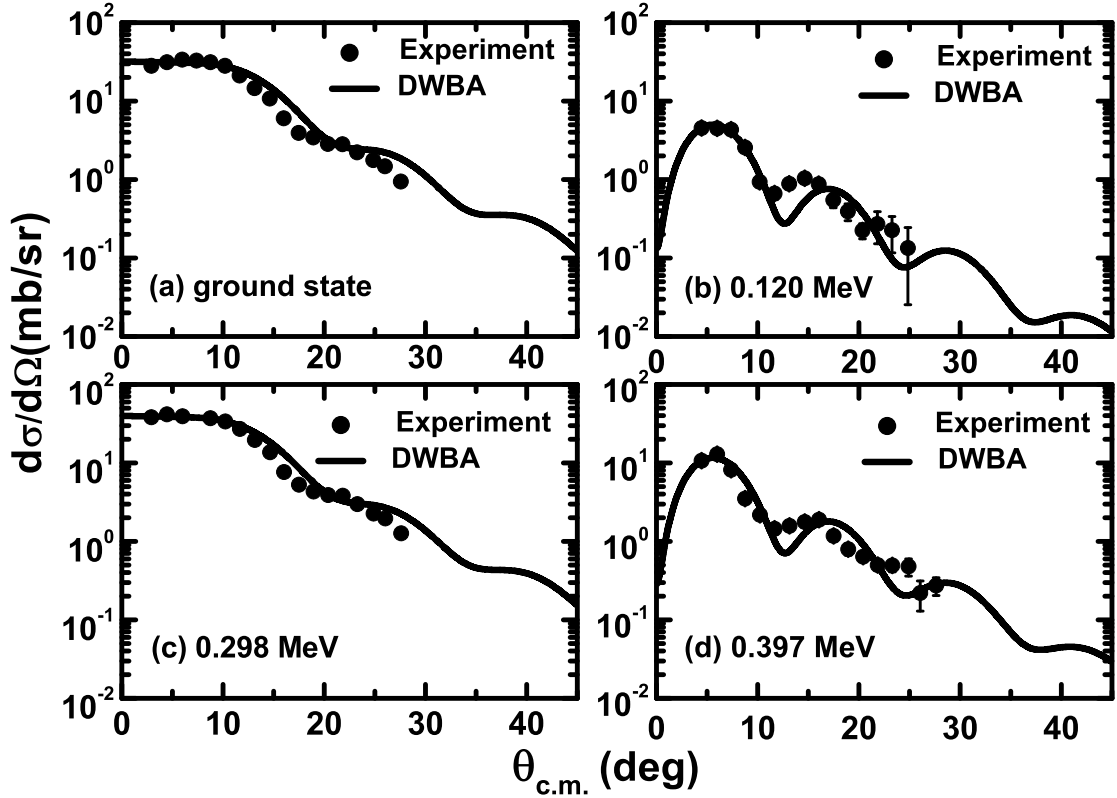}}
\caption{Angular distributions of the
$^{15}$N($^{7}$Li,\,$^{6}$Li)$^{16}$N reaction leading to the
ground and first three excited states in $^{16}$N. The curves
represent the DWBA calculations with the fitted OMP parameters.}
\label{fig5}
\end{center}
\end{figure*}

In Fig. \ref{fig4} we display the focal-plane position spectrum of
$^6$Li at $\theta_\mathrm{lab}$\,=\,10$^\circ$ from the
$^{15}$N($^{7}$Li,\,$^{6}$Li)$^{16}$N reaction leading to the
ground state and the first three excited states at $E_x$ = 0.120,
0.298, and 0.397 MeV in $^{16}$N. The closely spaced levels in
$^{16}$N were resolved and the background from $^{14}$N is
negligibly small. After background subtraction and beam
normalization, the angular distributions of the
$^{15}$N($^{7}$Li,\,$^{6}$Li)$^{16}$N reaction were obtained, as
presented in Fig. \ref{fig5}.

\section{Spectroscopic factors of the low-lying states in $^{16}$F}

The experimental angular distributions were analyzed with the
finite-range DWBA code FRESCO \cite{tho88}. The OMP parameters for
the entrance and exit channels were extracted by fitting the
present experimental angular distributions of the
$^{7}$Li\,+\,$^{15}$N and $^{6}$Li\,+\,$^{15}$N elastic scattering
(Fig. \ref{fig3}). The starting values of the OMP parameters were
obtained by fitting the systematic nucleus-nucleus potential based
on a single-folding model \cite{xu13}. The real potential was
chosen as a squared Woods-Saxon form, which fits the real part of
the folding model potential better than the usual Woods-Saxon form
does \cite{kho07}. For the imaginary potential the usual
Woods-Saxon form was found to be appropriate. In addition, we
investigated the effect of spin-orbit potential parameters
although for heavy ions they are thought to have little or no
influence on the cross sections \cite{tra00}. Full complex remnant
term interactions were included in the transfer reaction
calculations. The core-core ($^{6}$Li\,+\,$^{15}$N) potential
parameters were determined using the present ones of
$^{6}$Li\,+\,$^{15}$N at 34.5 MeV and the systematics in energy
dependence of the potential parameters of Ref. \cite{xu13}. For
the wave function of bound states, the Woods-Saxon potential with
the standard geometric parameters ($r$ = 1.25 fm and $a$ = 0.65
fm) was adopted, which have been extensively utilized to study the
ground state neutron spectroscopic factors for 80 nuclei of $Z$ =
3-24 \cite{tsa05} and 565 excited state neutron spectroscopic
factors for $Z$ = 8-28 nuclei \cite{tsa09}. The potential depths
were adjusted to reproduce the neutron binding energies. All the
parameters are listed in Table \ref{tab1}.

\begin{table*}
\begin{minipage}{1\textwidth}
\begin{center}
\caption{OMP parameters used in the present DWBA calculation.
$E_{\textrm{in}}$ denotes the incident energy in MeV for the
relevant channels, $V$ and $W$ are the depths (in MeV) of the real
and imaginary potentials with the squared Woods-Saxon form and the
usual Woods-Saxon form, and $r$ and $a$ are the radius and the
diffuseness (in fm). $\chi$$_\nu$$^{2}$ is the reduced chi-square
for the fitting. \label{tab1}}
\begin{tabular}{p{2cm}p{1cm}p{1.2cm}p{1cm}p{1cm}p{1cm}p{1cm}p{1cm}p{1cm}p{1cm}p{1cm}p{1cm}p{1cm}}
\hline\hline Channel & $E_{\textrm{in}}$ & $V$ & $r_{v}$ & $a_{v}$
& $W$ & $r_{w}$ & $a_{w}$& $V_{so}$ & $r_{so}$ &
$a_{so}$& $r_{C}$ & $\chi$$_\nu$$^{2}$\\
\hline
$^{7}$Li+$^{15}$N & 44.0 & 138.7 & 0.911 & 1.26 & 45.0 & 0.966 & 0.820 & & & & 1.30 & 4.08\\
$^{6}$Li+$^{16}$N & 34.5 & 111.0 & 0.886 & 1.47 & 39.0 & 0.840 & 1.02& & &  & 1.30 & 3.98\\
$^{6}$Li+$^{15}$N & 37.7 & 132.0 & 0.901 & 1.37 & 31.3 & 0.945 & 0.918& & &  & 1.30 & \\
$n$+$^{15}$N &  & ~~~\footnote{The depth was obtained by fitting
to reproduce the binding energy of the neutron in $^{16}$N.} & 1.25 & 0.65 & & & & 6.0 & 1.25 & 0.65 & 1.25 &\\
\hline\hline
\end{tabular}
\end{center}
\end{minipage}
\end{table*}

The spectroscopic factors of $^{16}$N can be derived by the
comparison of the experimental angular distribution with the DWBA
calculations using the relationship,
\begin{equation}\label{eq1}
    \sigma_{l,j}^{\textrm{exp}}(\theta)=
    S_{l,j}^{^{16}\textrm{N}}
    [S_{1,3/2}^{^7\textrm{Li}}\sigma_{1,3/2}^\textrm{DW}(\theta)+
    S_{1,1/2}^{^7\textrm{Li}}\sigma_{1,1/2}^\textrm{DW}(\theta)].
\end{equation}
Here $S_{l,j}^{^{16}\textrm{N}}$ is the spectroscopic factor of
$^{16}$N. $S_{1,3/2}^{^7\textrm{Li}}$ and
$S_{1,1/2}^{^7\textrm{Li}}$ are the spectroscopic factors of
$^7$Li, corresponding to the $j=3/2$ and $j=1/2$ orbits. The
square of the ANCs for the virtual decay
$^{16}$N\,$\rightarrow$\,$^{15}$N\,+\,$n$ was determined through
$(C_{l,j}^{^{16}\textrm{N}})^2 = S_{l,j}^{^{16}\textrm{N}} \times
(b_{l,j}^{^{16}\textrm{N}})^2$, where $b_{l,j}^{^{16}\textrm{N}}$
is the single-particle ANC of the bound state neutron in $^{16}$N.

To study $S_{l,j}^{^{16}\textrm{N}}$, $S_{1,3/2}^{^7\textrm{Li}}$
and $S_{1,1/2}^{^7\textrm{Li}}$ need to be determined. The value
of 0.73 was chosen as the total neutron spectroscopic factor
($S_{1,3/2}^{^7\textrm{Li}}$\,+\,$S_{1,1/2}^{^7\textrm{Li}}$) of
the $^7$Li ground state \cite{coh67,li69,tow69,su10}, as stated in
Ref. \cite{guo14}. According to the shell model calculation
\cite{coh67}, the ratio of $S_{1,3/2}^{^7\textrm{Li}}$ to
$S_{1,1/2}^{^7\textrm{Li}}$ was derived to be 1.5. The
spectroscopic factors of the ground state and the first three
excited states in $^{16}$N were then extracted to be
0.96\,$\pm$\,0.09, 0.69\,$\pm$\,0.09, 0.84\,$\pm$\,0.08 and
0.65\,$\pm$\,0.08, respectively. The errors result from the
statistics (8\%, 12\%, 8\%, 11\%), the uncertainty of target
thickness (5\%) and the uncertainty of spin-orbit potential
parameters (1.6\%, 2.2\%, 1.2\%, 3.1\%), respectively. The present
spectroscopic factors are approximately two times larger than
those from the $^{15}$N($d,\,p$) reaction \cite{boh72}, while they
are in good agreement with those from the $^2$H($^{15}$N,$\,p$)
reaction using Method2 (namely, components allowed to vary freely)
in Ref. \cite{bar08} where two different methods were used to
determine the spectroscopic factors since the closely spaced
levels (ground state + 0.120 MeV level, 0.298 + 0.397 MeV levels)
in $^{16}$N could not be resolved. It should be mentioned that the
relative spectroscopic factor values from all three measurements
agree within uncertainties. In addition, the squares of the ANCs
for the virtual decay $^{16}$N\,$\rightarrow$\,$^{15}$N\,+\,$n$
were derived to be 0.188\,$\pm$\,0.018, 3.54\,$\pm$\,0.46,
0.128\,$\pm$\,0.012 and 2.81\,$\pm$\,0.36 fm$^{\textrm{-1}}$,
respectively. All these results are listed in Table \ref{tab2}.

\begin{table}
\caption{Present spectroscopic factors of $^{16}$N and the square
of the ANCs for the virtual decay
$^{16}$N\,$\rightarrow$\,$^{15}$N\,+\,$n$. $nl_j$ is the
single-particle shell quantum number. \label{tab2}}
\begin{tabular}{p{2cm}p{1cm}p{1cm}cc}
\hline \hline $E_{x}$ (MeV)& $J^\pi$ & $nl_j$ & $S_{l,j}^{^{16}\textrm{N}}$ & $(C_{l,j}^{^{16}\textrm{N}})^2$ (fm$^{-1}$)\\
\hline
0 & 2$^-$ & 1$d_{5/2}$ & 0.96\,$\pm$\,0.09 & 0.188\,$\pm$\,0.018\\
0.120 & 0$^-$ & 2$s_{1/2}$ & 0.69\,$\pm$\,0.09 & 3.54\,$\pm$\,0.46\\
0.298 & 3$^-$ & 1$d_{5/2}$ & 0.84\,$\pm$\,0.08 & 0.128\,$\pm$\,0.012\\
0.397 & 1$^-$ & 2$s_{1/2}$ & 0.65\,$\pm$\,0.08 & 2.81\,$\pm$\,0.36\\
\hline \hline
\end{tabular}
\end{table}

We also investigated the dependence of the ANCs on the geometric
parameters of the Woods-Saxon potential for the single-particle
bound state in $^{16}$N. In the present calculation the radius was
adjusted and the new well depth was readjusted to reproduce the
binding energy. The result shows that for two levels corresponding
to neutron transfers to the 1$d_{5/2}$ orbit the spectroscopic
factors vary significantly, while the ANCs are nearly constant.
This indicates that the ANCs for these two levels are model
independent. Contrarily, the ANCs vary almost as significantly as
the spectroscopic factors do for two levels corresponding to
neutron transfers to the 2$s_{1/2}$ orbit, which indicates that
the ANCs for these two levels are model dependent. This difference
in response to transfers to the 1$d_{5/2}$ and 2$s_{1/2}$ states
may stem from the different peripheralities of these two
transitions.

\section{Proton widths of the low-lying resonant states in $^{16}$F}

The width $\Gamma_p$ of a proton resonance can be calculated
through
\begin{equation}\label{eq2}
    \Gamma_p = S_{l,j}^{^{16}\textrm{F}} \times \Gamma_p^{s.p.},
\end{equation}
where $\Gamma_p^{s.p.}$ denotes the single-particle width which
can be calculated from the scattering phase shift in a Woods-Saxon
potential. We assume that the spectroscopic factors for mirror
pair are equal
($S_{l,j}^{^{16}\textrm{F}}\,=\,S_{l,j}^{^{16}\textrm{N}}$)
according to charge symmetry of strong interaction, thus the
$\Gamma_p$ of $^{16}$F can be derived from the spectroscopic
factors of $^{16}$N via Eq. \ref{eq2}.

\begin{figure*}[htbp]
\begin{center}
\resizebox{0.8\textwidth}{!}{
  \includegraphics{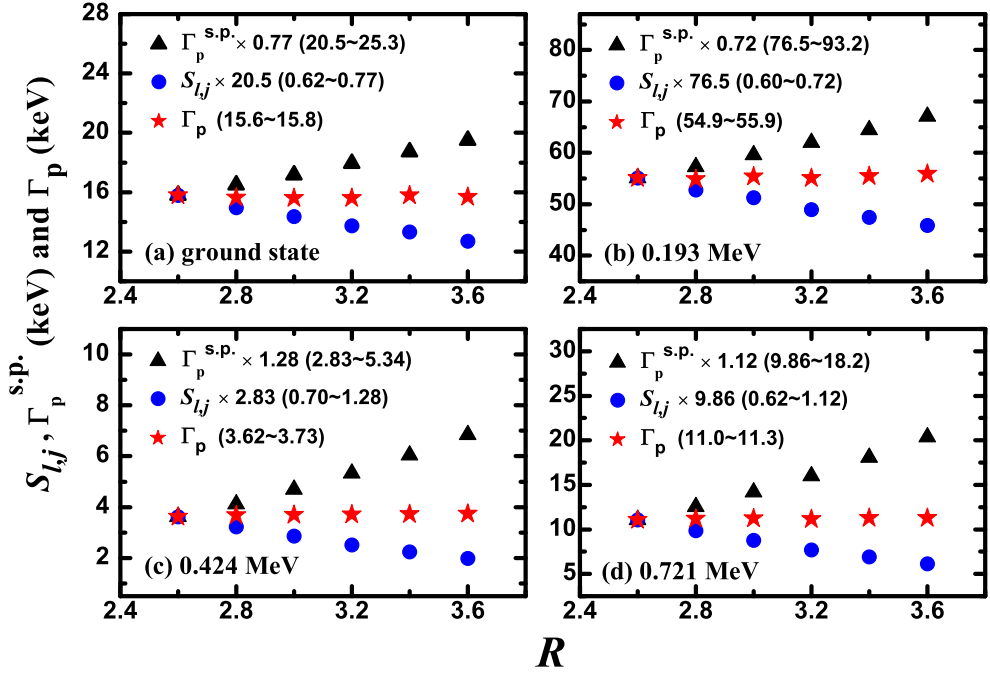}}
\caption{(Color online) Dependence of the single-particle width
($\Gamma_p^{s.p.}$), the spectroscopic factors of $^{16}$F
($S_{l,j}^{^{16}\textrm{F}}$), and the proton widths of $^{16}$F
($\Gamma_p$) on the radius ($R$). (a)-(d) represent the results
for the ground state and the first three excited states in
$^{16}$F, respectively. $\Gamma_p^{s.p.}$ and $S_{l,j}$ are
normalized to the $\Gamma_p$ value at $R = 2.6$ fm. The ranges in
the present results are given in parentheses.} \label{fig6}
\end{center}
\end{figure*}

We studied the dependence of the proton widths of $^{16}$F
($\Gamma_p$), the spectroscopic factors
($S_{l,j}^{^{16}\textrm{F}}$), and the single-particle width
($\Gamma_p^{s.p.}$) on the geometric parameter by changing the
radius ($R = r \cdot A^{1/3}$) within a reasonable range from 2.6
to 3.6 fm. The $^{16}$F spectroscopic factors are equal to the
$^{16}$N ones which were obtained using the new depths readjusted
to match the binding energies of the neutron in $^{16}$N. The
single-particle widths were computed with the new depths
determined by fitting to reproduce the resonance energies of the
proton in $^{16}$F. As shown in Fig. \ref{fig6}, the
single-particle widths and the spectroscopic factors of $^{16}$F
vary significantly, while the proton widths are nearly constant.
This indicates that the proton widths of the four $^{16}$F states
are model independent. The proton widths were derived to be
15.7\,$\pm$\,2.0, 55.3\,$\pm$\,7.2, 3.66\,$\pm$\,0.35, and
11.2\,$\pm$\,1.1 keV for these four states using the average
values for different radius, as listed in Table \ref{tab3}. The
uncertainties of geometric parameters were determined by taking
the half difference between the maximum and minimum widths in Fig.
\ref{fig6}. They were found to be less than 1.5\% for all four
levels in $^{16}$F, thus the error of the present proton widths
mainly results from the uncertainty of the spectroscopic factors.

\begin{table}
%\begin{minipage}{6.5cm}
\begin{center}
\caption{The spectroscopic factors ($S_{l,j}^{^{16}\textrm{F}}$),
the single-particle width ($\Gamma_p^{s.p.}$), and the proton
widths of $^{16}$F ($\Gamma_p$). $S_{l,j}^{^{16}\textrm{F}}$ and
$\Gamma^{s.p.}$ are obtained with standard geometric parameters,
while $\Gamma_p$ are the average values for the radius range from
$R$ = 2.6 to 3.6 fm \label{tab3}}
\begin{tabular}{p{2cm}p{1cm}p{1.5cm}ccc}
\hline \hline $E_{x}$(MeV) & $J^\pi$ & $nl_j$ &
$S_{l,j}^{^{16}\textrm{F}}$~ & $\Gamma^{s.p.}$ (keV)~ & $\Gamma_p$
(keV)\\
\hline
0.000 & 0$^-$ & 2$s_{1/2}$ & 0.69\,$\pm$\,0.09 & 22.7 & 15.7\,$\pm$\,2.0\\
0.193 & 1$^-$ & 2$s_{1/2}$ & 0.65\,$\pm$\,0.08 & 84.1 & 55.3\,$\pm$\,7.2\\
0.424 & 2$^-$ & 1$d_{5/2}$ & 0.96\,$\pm$\,0.09 & 3.86 & 3.66\,$\pm$\,0.35\\
0.721 & 3$^-$ & 1$d_{5/2}$ & 0.84\,$\pm$\,0.08 & 13.3 & 11.2\,$\pm$\,1.1\\
\hline \hline
\end{tabular}
\end{center}
%\end{minipage}
\end{table}

\begin{table*}
%\begin{minipage}{16cm}
\begin{center}
\caption{Present $^{16}$F proton widths in keV and other available
results in the literature. \label{tab4}}
\begin{tabular}{cccccccc}
\hline \hline $E_{x}$ & Compilation&$^{14}$N($^3$He,\,$n$) &
$^{14}$N($^3$He,\,$np$) & $^{16}$O($^3$He,\,$t$) &
$^{16}$O($^3$He,\,$t$) &
$p$($^{15}$O,\,$p$) & $^{15}$N($^{7}$Li,\,$^{6}$Li)\\
(MeV)&\cite{til93}&\cite{zaf65}&\cite{ots76}&\cite{ste84}&\cite{fuj09}&\cite{lee07}& Present\\
\hline
0.000 & 40\,$\pm$\,20 & 50\,$\pm$\,30 & 39\,$\pm$\,20 & $\approx$\,25 & 18\,$\pm$\,16 & 22.8\,$\pm$\,7.2 & 15.7\,$\pm$\,2.0\\
0.193 & $<$\,40 & $<$\,40 & 96\,$\pm$\,20 & $\approx$\,100 & 87\,$\pm$\,16 & 103\,$\pm$\,6 & 55.3\,$\pm$\,7.2\\
0.424 & 40\,$\pm$\,30 & 40\,$\pm$\,30 & 24\,$\pm$\,20 && 16\,$\pm$\,16 & 4.0\,$\pm$\,1.3 & 3.66\,$\pm$\,0.35\\
0.721 & $<$\,15 & $<$\,15 & 24\,$\pm$\,20 && 12\,$\pm$\,16 & 15.1\,$\pm$\,3.4 & 11.2\,$\pm$\,1.1\\
\hline \hline
\end{tabular}
\end{center}
%\end{minipage}
\end{table*}

In Table \ref{tab4} we compare different evaluations of the proton
widths from the present work and the previous studies. The present
width of the $^{16}$F ground state is narrower than the lower
limits from the compilation \cite{til93} and the
$^{14}$N($^3$He,\,$n$) data \cite{zaf65,ots76}, and is narrower
than the value from the $^{16}$O($^3$He,\,$t$) data \cite{ste84}.
The new width of the first excited state is larger than the upper
limits of Refs. \cite{til93,zaf65}, while is narrower than those
of Refs. \cite{ots76,ste84,fuj09}. The present width of the second
excited state is narrower than the lower limits from the
compilation \cite{til93} and the $^{14}$N($^3$He,\,$n$) data
\cite{zaf65,ots76}. In addition, our results are in good agreement
with those from the most recent $p$($^{15}$O,\,$p$) data
\cite{lee07} for all the levels except the first excited state.
The width of 103\,$\pm$\,9 keV for the first excited state given
in Ref. \cite{lee07} would yield a spectroscopic factor of
1.22\,$\pm$\,0.11, which is significantly larger than the present
result (0.65\,$\pm$\,0.08) from the
$^{15}$N($^{7}$Li,\,$^{6}$Li)$^{16}$N data and that
(0.74\,$\pm$\,0.12) from the $^{15}$N($d$,$\,p$)$^{16}$N data
\cite{bar08} and the shell model prediction (0.96) \cite{mei96}.
Therefore, additional measurements of this width via an
independent method are certainly desirable.

\section{Discussion and conclusion}

The angular distributions of the
$^{15}$N($^{7}$Li,\,$^{6}$Li)$^{16}$N reaction were measured by a
high-precision Q3D magnetic spectrograph and were utilized to
determine the neutron spectroscopic factors and the ANCs for the
four low-lying $^{16}$N states. We also investigated the
dependence of our results on the geometric parameters of the
Woods-Saxon potential for the single-particle bound state in
$^{16}$N. It was found that the ANCs for the two levels
corresponding to neutron transfers to the 1$d_{5/2}$ orbit are
more model independent than the ANCs for the two levels
corresponding to neutron transfers to the 2$s_{1/2}$ orbit. This
difference may come from the different peripheralities of these
two transitions.

The proton widths of the four low-lying levels in $^{16}$F were
determined from the $^{16}$N spectroscopic factors by charge
symmetry of mirror nuclei. In addition, we studied the dependence
of the proton widths on the geometric parameters of the
Woods-Saxon potential. The result demonstrates that the proton
widths of these four states in $^{16}$F are all model independent.
The new widths are in good agreement with those from the most
recent $p$($^{15}$O,\,$p$) data \cite{lee07} for the ground state,
the second and third excited states in $^{16}$F. For the first
excited state the present width is nearly half of that in Ref.
\cite{lee07}. To understand this discrepancy additional
measurements of this width via an independent method are highly
desirable.

\begin{acknowledgments}

We acknowledge the staff of HI-13 tandem accelerator for the
smooth operation of the machine, and N. K. Timofeyuk and R. C.
Johnson for their helpful discussions on DWBA calculations and
charge symmetry. This work was performed with the support of the
National Natural Science Foundation of China under Grant Nos.
11321064, 11075219, 11375269, 11275272 and 11275018, the 973
program of China under Grant No. 2013CB834406.

\end{acknowledgments}

%\end{CJK*}


\begin{thebibliography}{}

\bibitem{zaf65} {C. D. Zafiratos, F. Ajzenberg-Selove and F. S. Dietrich, Phys. Rev. {\bf 137}, B1479 (1965).}
\bibitem{boh73} {W. Bohne, H. Fuchs, K. Grabisch et al., Phys. Lett. B {\bf 47}, 342 (1973).}
\bibitem{ots76} {T. Otsubo, I. Asada, M. Takeda et al., Nucl. Phys. A {\bf 259}, 452 (1976).}
\bibitem{mos71} {C. E. Moss and A. B. Comiter, Nucl. Phys. A {\bf 178}, 241 (1971).}
\bibitem{faz82} {A. Fazely, B. D. Anderson, M. Ahmad et al., Phys. Rev. C {\bf 25}, 1760 (1982).}
\bibitem{ori82} {H. Orihara, S. Nishihara, K. Furukawa et al., Phys. Rev. Lett. {\bf 49}, 1318 (1982).}
\bibitem{ohn87} {H. Ohnuma, M. Kabasawa, K. Furukawa et al., Nucl. Phys. A {\bf 467}, 61 (1987).}
\bibitem{mad97} {R. Madey, B. S. Flanders, B. D. Anderson et al., Phys. Rev. C {\bf 56}, 3210 (1997).}
\bibitem{peh65} {R. H. Pehl and J. Cerny, Phys. Lett. {\bf 14}, 137 (1965).}
\bibitem{nan77} {H. Nann, W. Benenson, E. Kashy, H. P. Morsch and D. Mueller, Phys. Rev. C {\bf 16}, 1684 (1977).}
\bibitem{ste84} {W. A. Sterrenburg, S. Brandenburg, J. H. Van Dijk et al., Nucl. Phys. A {\bf 420}, 257 (1984).}
\bibitem{fuj09} {H. Fujita, G. P. A. Berg, Y. Fujita et al., Phys. Rev. C {\bf 79}, 024314 (2009).}
\bibitem{til93} {D. R. Tilley, H. R. Weller and C. M. Cheves, Nucl. Phys. A {\bf 564}, 1 (1993).}
\bibitem{lee07} {D. W. Lee, K. Per$\ddot{\textrm{a}}$j$\ddot{\textrm{a}}$rvi, J. Powell et al., Phys. Rev. C {\bf 76}, 024314 (2007).}
\bibitem{guo07} {B. Guo, Z. H. Li, W. P. Liu et al., J. Phys. G {\bf 34}, 103 (2007).}
\bibitem{tim06} {N. K. Timofeyuk, D. Baye, P. Descouvemont, R. Kamouni and I. J. Thompson, Phys. Rev. Lett. {\bf 96}, 162501 (2006).}
\bibitem{guo06} {B. Guo, Z. H. Li, X. X. Bai et al., Phys. Rev. C {\bf 73}, 048801 (2006).}
\bibitem{reh98} {K. E. Rehm, F. Borasi, C. L. Jiang et
al., Phys. Rev. Lett. {\bf 80}, 676 (1998).}
\bibitem{guo14} {B. Guo, Z. H. Li, Y. J. Li et al., Phys. Rev. C {\bf 89}, 012801(R) (2014).}
\bibitem{guo12} {B. Guo, Z. H. Li, M. Lugaro et al., Astrophys. J. {\bf 756}, 193 (2012).}
\bibitem{li12} {Y. J. Li, Z. H. Li, E. T. Li et al., Eur. Phys. J. A {\bf 48}, 13 (2012).}
\bibitem{li13} {Z. H. Li, Y. J. Li, J. Su et al., Phys. Rev. C {\bf 87}, 017601 (2013).}
\bibitem{woo82} {C. L. Woods, B. A. Brown and N. A. Jelley, J. Phys. G {\bf 8}, 1699 (1982).}
\bibitem{oli96} {F. de Oliveira, A. Coc, P. Aguer et al., Nucl. Phys. A {\bf 597}, 231 (1996).}
\bibitem{tho88} {I. J. Thompson, Comput. Phys. Rep. {\bf 7}, 167 (1988).}
\bibitem{xu13} {Y. P. Xu and D. Y. Pang, Phys. Rev. C {\bf 87}, 044605 (2013).}
\bibitem{kho07} {D. T. Khoa, W. V. Oertzen, H. G. Bohlen and S. Ohkubo, J. Phys. G {\bf 34}, R111 (2007).}
\bibitem{tra00} {L. Trache, A. Azhari, H. L. Clark et al., Phys. Rev. C {\bf 61},
024612 (2000).}
\bibitem{tsa05} {M. B. Tsang, J. Lee and W. G. Lynch, Phys. Rev. Lett. {\bf 95},
222501 (2005).}
\bibitem{tsa09} {M. B. Tsang, J. Lee, S. C. Su et al., Phys. Rev. Lett. {\bf 102},
062501 (2009).}
\bibitem{coh67} {S. Cohen and D. Kurath, Nucl. Phys. A {\bf 101}, 1 (1967).}
\bibitem{li69} {T. Y. Li and S. K. Mark, Nucl. Phys. A {\bf 123}, 147 (1969).}
\bibitem{tow69} {I. S. Towner, Nucl. Phys. A {\bf 126}, 97 (1969).}
\bibitem{su10} {J. Su, Z. H. Li, B. Guo et al., Chin. Phys. Lett. {\bf 27}, 052101 (2010).}
\bibitem{boh72} {W. Bohne, J. Bommer, H. Fuchs et al., Nucl. Phys. A {\bf 196}, 41
(1972).}
\bibitem{bar08} {D. W. Bardayan, P. D. O'Malley, J. C. Blackmon et al., Phys. Rev. C {\bf 78}, 052801(R) (2008).}
\bibitem{mei96} {J. Meissner, H. Schatz, H. Herndl et al., Phys. Rev. C {\bf 53}, 977 (1996).}

\end{thebibliography}
\end{document}